\pdfoutput=1
\documentclass{article}

\usepackage[final]{neurips_2023}

\usepackage[utf8]{inputenc} 
\usepackage[T1]{fontenc}    
\usepackage[hidelinks]{hyperref} 
\usepackage{url}            
\usepackage{booktabs}       
\usepackage{amsfonts}       
\usepackage{nicefrac}       
\usepackage{microtype}      
\usepackage{xcolor}         
\usepackage[hang,flushmargin]{footmisc}
\usepackage{enumitem}
\usepackage{xspace}
\usepackage{arydshln}
\usepackage{multirow}
\usepackage{graphicx}

\usepackage{etoolbox}
\makeatletter
\patchcmd{\@verbatim}
  {\verbatim@font}
  {\verbatim@font\small}
  {}{}
\makeatother

\newcommand{\ignore}[1]{}
\newcommand{\mytt}[1]{\texttt{\small #1}}

\title{Searching Dense Representations\\ with Inverted Indexes}

\author{Jimmy Lin$^{1}$ \textnormal{and} Tommaso Teofili$^2$ \\[1ex]
$^{1}$ David R.\ Cheriton School of Computer Science, University of Waterloo\\
$^{2}$ Department of Engineering, Roma Tre University
}

\begin{document}
\maketitle

\begin{abstract}
Nearly all implementations of top-$k$ retrieval with dense vector representations today take advantage of hierarchical navigable small-world network (HNSW) indexes.
However, the generation of vector representations and efficiently searching large collections of vectors are distinct challenges that can be decoupled.
In this work, we explore the contrarian approach of performing top-$k$ retrieval on dense vector representations {\it using inverted indexes}.
We present experiments on the MS MARCO passage ranking dataset, evaluating three dimensions of interest:\ output quality, speed, and index size.
Results show that searching dense representations using inverted indexes {\it is} possible.
Our approach exhibits reasonable effectiveness with compact indexes, but is impractically slow.
Thus, while workable, our solution does not provide a compelling tradeoff and is perhaps best characterized today as a ``technical curiosity''.
\end{abstract}

\section{Introduction}

In so-called dense retrieval models~\citep{karpukhin-etal-2020-dense}, queries and passages are both encoded as dense vector representations (often called embeddings), and top-$k$ retrieval is formulated as a nearest neighbour search problem.
That is, given a query vector, the system's task is to retrieve the top-$k$ most similar passage vectors with respect to a simple comparison operation, typically the inner (dot) product.
Today, these dense representation vectors are typically derived from transformer-based models fine-tuned on a dataset of relevant query--passage pairs; a common configuration involves models that generate vectors of 768 dimensions.

Despite the observation that dense retrieval models and sparse bag-of-words lexical models such as BM25 capture parametric variations of a bi-encoder architecture~\citep{Lin_arXiv2021_repir}, implementations of top-$k$ retrieval are quite different for the two classes of models.
For sparse bag-of-words vectors, the venerable inverted index has served as the workhorse for top-$k$ retrieval dating back many decades.
For dense vectors, current best practices take advantage of hierarchical navigable small-world network (HNSW) indexes~\citep{HNSW} to perform approximate nearest neighbour search; the Faiss~\citep{faiss} library provides an implementation that is widely used today.

\citet{Lin_arXiv2021_repir} pointed out that the core retrieval problem can be decomposed into two independent components, what he refers to as the logical scoring model and the physical retrieval model.
That is, the generation of vector representations from content is distinct from efficient solutions to the top-$k$ retrieval problem.
Of course, there exists a strong affinity between sparse representations and inverted indexes, on the one hand, and dense representations and HNSW indexes, on the other.
However, this tight coupling does not necessarily need to be the case.

In this work, we explore the contrarian approach of searching dense representations with inverted indexes.
Building on previous work~\citep{Teofili_Lin_arXiv2019}, we apply two types of transformations---``fake words'' and ``lexical LSH''---that enable dense representations to be captured in standard inverted indexes and we empirically evaluate top-$k$ retrieval using these two techniques.
Such a solution is potentially interesting because it enables dense and sparse retrieval using a single infrastructural component, obviating the need to maintain and coordinate different types of indexes.

We present experiments on the MS MARCO passage ranking dataset, evaluating three dimensions of interest:\ output quality, speed, and index size.
Results show that it {\it is} possible to perform top-$k$ retrieval on dense representations using only inverted indexes:\
Compared to HNSW indexes, we can achieve reasonable effectiveness with much smaller indexes, but unfortunately, search is impractically slow.
Thus, while our proposed techniques are ``workable'', they do not appear to provide a compelling tradeoff in the overall design space.
We would characterize them as a ``technical curiosity'', but still worthwhile for the community to be aware of.
Perhaps our efforts will become a part of further breakthroughs that will yield a practical solution.

\section{Methods}

In this work, we examined two techniques for top-$k$ retrieval on dense vectors using inverted indexes.
Both techniques were originally implemented in the Anserini toolkit~\citep{anserini} using ``stock parts'' from the open-source Lucene search library, as part of the work described in~\cite{Teofili_Lin_arXiv2019}.
However, this previous work pre-dated the advent of dense retrieval models and focused on similarity comparisons with word embeddings, which lacked a concrete task.
Here, we applied the same techniques, but to an actual real-world retrieval application.

\paragraph{``Fake words''.}
We implemented the approach described in~\cite{Amato_etal_2016}, which encodes the features of a vector as a number of ``fake'' terms proportional to the feature value according to the following scheme:

Given a vector $w = (w_1, \ldots , w_m)$, each feature $w_i$ is associated with a unique alphanumeric term $\tau_i$ such that the document corresponding to the vector $w$ is represented by ``fake words'' generated by
$$\cup_{i=1}^{m}\cup_{j=1}^{\lfloor Q \cdot w_i \rfloor}\tau_i,$$ 
where $Q>1$ is a quantization factor.
Thus, the fake words encoding maintains direct proportionality between the float value of a feature and the term frequency of the corresponding fake index term.
Feature-level matching for retrieval is achieved by matching on these fake words with scores computed by Lucene's \mytt{ClassicSimilarity}, which is a tf---idf variant.
Finally, for this approach to be effective, vector inner products have to be equivalent to cosine similarity, which can be achieved by normalizing the vectors to unit length.

\paragraph{``Lexical LSH''.} 
We implemented an approach that \textit{lexically} quantizes vector components for easy indexing and search in Lucene using LSH~\citep{gionis1999similarity}.
While LSH is, of course, not new, to our knowledge, \cite{lexlsh} was the first to devise an implementation that directly integrates with inverted indexes inside Lucene.

Given a vector $w = (w_1, \ldots , w_m)$, each feature $w_i$ is rounded to the $d$-th decimal place and tagged with its feature index $i$ as a term prefix.
For example, consider $w = \{0.12, 0.43, 0.74\}$.
If $d=1$, $w$ is converted into the tokens \mytt{1\_0.1}, \mytt{$2\_0.4$}, and \mytt{$3\_0.7$}.
In our implementation, tokens are aggregated into $n$-grams and finally passed to an LSH function, which is implemented in Lucene as \mytt{MinHashFilter}, to hash the $n$-grams into a configurable number of buckets $b$.
Thus, the vector $w$ is represented as a set of LSH-generated text signatures for \textit{tagged} and \textit{quantized} feature $n$-grams.

\section{Experiments}

While our techniques are agnostic with respect to the actual dense retrieval model, for fair comparisons to HNSW indexes in Lucene, we needed vector representations that are normalized to unit length because Lucene's implementation is restricted to top-$k$ retrieval using cosine similarity (as opposed to general inner products).
Many dense retrieval models generate representations that do not perform this normalization.
For their HNSW experiments in Lucene, \citet{Ma_etal_CIKM2023} had to fine-tune a new embedding model from scratch, which they called cosDPR-distil.
To facilitate comparisons to this work, we used the same model.

We evaluated top-$k$ retrieval using standard evaluation methodology on the MS MARCO passage ranking test collection~\citep{msmarco}, comprising three separate sets of queries:\ the 6980 queries from the development (dev) set, as well as queries from the TREC 2019 and 2020 Deep Learning Tracks~\citep{trec2019,trec2020}.

Our experiments were performed with Anserini at commit \mytt{e99c73d} (11/25/2023) on a Mac Studio with an M1 Ultra processor containing 20 cores (16 performance and 4 efficiency) and 128 GB memory, running macOS Sonoma 14.1.1 and OpenJDK 11.0.13.
Unless otherwise specified, all runs used 16 threads.

\begin{table}[t]
\centering
\scalebox{0.85}{
\setlength{\tabcolsep}{3pt}
\begin{tabular}{lccrccccr}
\toprule
& \multicolumn{2}{c}{{\bf dev}} & & \multicolumn{2}{c}{{\bf DL19}} & \multicolumn{2}{c}{{\bf DL20}} & Index Size \\
& RR@10 & R@1k & QPS & nDCG@10 & R@1k & nDCG@10 & R@1k & (GB) \\
\cmidrule(lr){2-4} \cmidrule(lr){5-6} \cmidrule(lr){7-8} 
BM25 & 0.1840 & 0.8526 & 426.49 & 0.5058 & 0.7501 & 0.4796 & 0.7863 & 2.5  \\
\midrule
FW ($Q=10$) & 0.0045 & 0.0440 & 940.36 & 0.0063 & 0.0167 & 0.0241 & 0.0408 & 0.2 \\
FW ($Q=20$) & 0.2937 & 0.9142 & 18.65 & 0.5795 & 0.7238 & 0.5912 & 0.7327 & 1.3 \\
FW ($Q=30$) & 0.3498 & 0.9580 & 5.31 & 0.6488 & 0.7788 & 0.6483 & 0.7980 & 2.8 \\
FW ($Q=40$) & 0.3605 & 0.9668 & 2.80 & 0.6857 & 0.7902 & 0.6666 & 0.8194 & 4.2 \\
FW ($Q=50$) & 0.3627 & 0.9669 & 1.91 & 0.6930 & 0.7957 & 0.6724 & 0.8193 & 5.7 \\
FW ($Q=60$) & 0.3681 & 0.9707 & 1.56 & 0.6849 & 0.8005 & 0.6832 & 0.8261 & 7.1 \\
FW ($Q=70$) & 0.3657 & 0.9695 & 1.42 & 0.6933 & 0.7979 & 0.6823 & 0.8223 & 8.4 \\
FW ($Q=80$) & 0.3642 & 0.9708 & 1.25 & 0.6833 & 0.8015 & 0.6706 & 0.8267 & 9.7 \\
FW ($Q=90$) & 0.3668 & 0.9733 & 1.19 & 0.7013 & 0.8006 & 0.6750 & 0.8271 & 11 \\
\midrule
LexLSH ($b=100$) & 0.2284 & 0.8365 & 6.26 & 0.4233 & 0.5614 & 0.4880 & 0.6391 & 0.9 \\
LexLSH ($b=200$) & 0.2959 & 0.9267 & 3.10 & 0.5810 & 0.6610 & 0.5863 & 0.7485 & 1.4 \\
LexLSH ($b=300$) & 0.3180 & 0.9457 & 2.06 & 0.6167 & 0.7272 & 0.6265 & 0.7866 & 2.1 \\
LexLSH ($b=400$) & 0.3309 & 0.9538 & 1.41 & 0.6398 & 0.7450 & 0.6505 & 0.7924 & 2.7 \\
LexLSH ($b=500$) & 0.3397 & 0.9569 & 1.09 & 0.6443 & 0.7556 & 0.6548 & 0.7992 & 3.3 \\
LexLSH ($b=600$) & 0.3457 & 0.9596 & 0.83 & 0.6716 & 0.7610 & 0.6569 & 0.8131 & 3.9 \\
LexLSH ($b=700$) & 0.3474 & 0.9609 & 0.74 & 0.6843 & 0.7735 & 0.6558 & 0.8157 & 4.5 \\
LexLSH ($b=800$) & 0.3496 & 0.9611 & 0.67 & 0.6778 & 0.7784 & 0.6669 & 0.8181 & 4.9 \\
LexLSH ($b=900$) & 0.3496 & 0.9611 & 0.66 & 0.6778 & 0.7784 & 0.6669 & 0.8181 & 4.9 \\
\midrule
HNSW (default)  & 0.3881 & 0.9732 & 47.78 & 0.7159 & 0.8101 & 0.6967 & 0.8391 & 26 \\
HNSW (optimized) & 0.3885 & 0.9747 & 387.29 & 0.7250 & 0.8222 & 0.7025 & 0.8520 & 26 \\
\bottomrule
\end{tabular}
}
\vspace{0.3cm}
\caption{Performance of our proposed fake words and lexical LSH techniques on the MS MARCO passage corpus.}
\label{results}
\end{table}

Results are presented in Table~\ref{results}, covering the aspects of performance that we are interested in:\
output quality as measured in standard IR effectiveness metrics, speed in terms of query throughput (measured in queries per second or QPS), and index size (measured with the \mytt{du -h} command).
The rows capture results either with the fake words technique (FW), parameterized by $Q$, or the lexical LSH technique, parameterized by the number of buckets $b$ (with $d=1$).
Query throughput is measured only on the dev set, which has a sufficient number of queries (6980) to obtain reliable measurements; we observe only small variations from run to run.
We report the average of three trials.
In all cases, experimental runs used pre-encoded queries---that is, cached representations from neural inference applied to the queries.
To better understand the overhead associated with query inference, we refer readers to evaluations in~\citet{Chen_etal_arXiv2023}.

For reference, evaluation of BM25 is presented in the top row, and evaluation of HNSW indexes for cosDPR-distil is presented in the final two rows.
The ``default'' HNSW condition characterizes performance using Anserini ``out of the box'' with default parameters (\mytt{M} set to 16, \mytt{efC} set to 100, 16 indexing threads).
The ``optimized'' index was constructed with \mytt{efC} set to 1000 (all other parameters being the same), but optimized down to a single index segment (which is a very time-consuming operation); this is the same exact index instance used in~\citet{Chen_etal_arXiv2023}.
This optimization greatly increases search performance, but unless the document collection is static, this step is unrealistic for real-world use.
Note that since HNSW indexing is non-deterministic, different index instances (i.e., from running the indexer multiple times)\ may exhibit small effectiveness variations.
We report scores from our specific index instances, which may differ slightly from the official Anserini reproducibility documentation.

From Table~\ref{results}, looking at the fake words technique, it appears that the sweet spot is around $Q=40$.
A bit higher effectiveness comes at a roughly 30$\%$ decrease in QPS, and the index size of 2.8 GB remains quite modest.
For the lexical LSH technique, the sweet spot appears to be around $b=400$, with a 2.7 GB index.
At a high level, it appears that the fake words technique provides better tradeoffs than the lexical LSH technique.
Overall, for both techniques we would characterize the effectiveness as ``acceptable'', with compact inverted indexes that are much smaller than the HNSW indexes.
However, search is impractically slow compared to retrieval based on HNSW indexes.

While it would be possible to perform more exhaustive parameter tuning, better parameter selection alone is unlikely to close the performance gap between either technique and HNSW indexes.
The much smaller index sizes offered by our techniques definitely present an advantage over HNSW indexes, but we do not see a compelling use case for either of these techniques.
Thus, we would characterize the techniques presented here as ``technical curiosities'', but impractical overall.
Although we can imagine a number of further explorations, such as hybrid dense--sparse models with inverted indexes, further pursuit of these avenues does not seem particularly promising at present.
It is clear that more breakthroughs are needed, and perhaps this work will represent a step along the way.

\section{Conclusions}

This work demonstrates that it {\it is} indeed possible to perform top-$k$ retrieval with dense vector representations {\it using only inverted indexes} in the popular Lucene search toolkit.
This contrarian approach eliminates the need for dedicated infrastructure such as vector stores and HNSW indexes.
Of the two techniques we examined, the fake words technique appears to represent a better solution.
In terms of the three dimensions of performance that we are interested in, the technique excels at optimizing for storage, with indexes that are far smaller than HNSW indexes.
However, search is impractically slow compared to the HNSW index implementation in Lucene, and we cannot at present imagine a scenario where this solution is compelling.

\section*{Acknowledgements}

This research was supported in part by the Natural Sciences and Engineering Research Council (NSERC) of Canada.

\bibliography{custom}
\bibliographystyle{abbrvnat}

\end{document}